\documentclass[aps,pre, reprint, showpacs]{revtex4-1}
\usepackage{amstext,amsmath}
\usepackage{graphicx}
\usepackage{color}
\usepackage[normalem]{ulem}

\usepackage[dvipsnames]{xcolor}

\begin{document}

\title{Driven transport on a flexible polymer with particle recycling: a model inspired by transcription and translation}
\author{Lucas D.~Fernandes}
\affiliation{Departamento de Entomologia e Acarologia, Escola Superior de Agricultura Luiz de Queiroz - Universidade de S\~{a}o Paulo (USP), 13418-900, Piracicaba/SP, Brazil}
\affiliation{Department of Life Sciences, Imperial College London, Silwood Park, Ascot, Berkshire, SL5 7PY, UK}
\author{Luca Ciandrini}
\email{luca.ciandrini@umontpellier.fr}
\affiliation{Laboratoire Charles Coulomb (L2C), Universit\'e de Montpellier and CNRS, Montpellier, France}
\affiliation{CBS, Universit\'e de Montpellier, CNRS and INSERM, Montpellier, France}
\affiliation{DIMNP, Universit\'e de Montpellier and CNRS, Montpellier, France}

\date{4 May 2019}

\begin{abstract}
Many theoretical works have attempted to coarse grain gene expression at the level of transcription and translation via frameworks based on exclusion processes. Usually in these models the three-dimensional conformation of the substrates (DNA and mRNA) is neglected, and particles move on a static unidimensional lattice in contact to an infinite reservoir. In this work we generalise the paradigmatic exclusion process and study the transport of particles along a unidimensional polymer-like flexible lattice immersed in a three-dimensional particle reservoir. We study the recycling of particles in the reservoir, how the transport is influenced by the global conformation of the lattice and, in turn, how particle density dictates the structure of the polymer.
\end{abstract}
	
\maketitle
\section{Introduction} 
Since their first formulation at the end of the 60's~\cite{*MacDonald1968KineticsTemplates,*MacDonald1969ConcerningPolyribosomes}, driven lattice models drew the attention of the scientific community for both their relevance in non-equilibrium statistical mechanics, the novelty of their theoretical approaches, and their powerful applications in transport processes~\cite{Chou2011Non-equilibriumTransport}.

Historically, the prototypical model of unidimensional traffic, the totally asymmetric simple exclusion process (TASEP), has been developed and then extended to describe the collective movement of biological ``active particles'' such as ribosomes, RNA polymerases or motor proteins, on their respective unidimensional substrates (mRNA, DNA, microtubules or actin filaments). Most of state-of-the-art models describing the gene expression stages of transcription and translation exploit implicitly or explicitly this class of models~\cite{Mitarai2008RibosomeDestabilization,Reuveni2011Genome-ScaleModel,Leduc2012MolecularMicrotubules., Ciandrini2013RibosomeRegulation}; despite their coarse-grained nature, these frameworks are able to capture the main features of the biological processes.

Although it is incontestable that strands of mRNA or DNA molecules are dynamical objects with complex three-dimensional conformations, common models approximate those tracks with unidimensional unstructured lattices and neglect their polymer-like nature. The interdependence between the lattice conformation and the transport process, however, should be considered when focusing on quantitative modelling aimed to compare experimental data and extract information on the molecular mechanisms. 
For instance, spatial clustering of genes in transcription factories~\cite{Sutherland2009TranscriptionUnions} suggests an interplay between structural conformation of the DNA, gene expression and local recycling of polymerases. Similar effects, including the importance of local ribosome concentrations, can also be expected in translation since the ribosomes are not uniformly distributed in the cytoplasm~\cite{Bakshi2012SuperresolutionCells}. Furthermore, different conformations assumed by the transcript can explain the gene length-dependence of mRNA translation, as we have recently addressed~\cite{Fernandes2017}. Although there are a few models considering the effects of the transport on the substrate dynamics~\cite{Melbinger2012MicrotubuleMotors} or on local structures~\cite{VonHeijne1977TranslationStructure,*Turci2013TransportDefects}, its impact on the overall three-dimensional conformation of the lattice has not been explored.

In this work we propose a non-equilibrium model of transport on a polymer-like substrate, which is immersed in a three-dimensional reservoir of diffusing particles. In our derivation we implicitly consider that the timescales of transport and polymer dynamics are well separated~\cite{Al-Hashimi2008RNATime.}: polymerases or ribosomes move at a speed of $\sim 10$ nm/s, while the dynamics of structural elements of nucleotide chains is orders of magnitude faster~\cite{Al-Hashimi2008RNATime.}. In case of mRNA, for instance, the ribosomes advances roughly 10 persistence lengths of the polymer chain (which is roughly $10\times1$ nm) in a second. Hence, it is reasonable to assume that the polymer can equilibrate after each particle step, and the local polymer conformation does not affect the motion of particles. We thoroughly explain the assumptions of the model in Appendix~\ref{app:approx}.

We investigate (i) how the three-dimensional structure of the lattice affects the particle recruitment and the transport process, as well as (ii) how the driven lattice gas impacts, via the particle density, global features of the polymer. We start with a short review of the well known TASEP results, then we couple the system to a three-dimensional diffusive reservoir of particles, and eventually study the interplay between the local concentration of particles, the lattice conformation and the transport process.

\begin{figure}[t]
	\centering
	\includegraphics[width=.95\columnwidth]{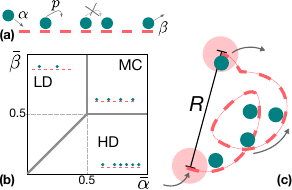}  
	\caption{Scheme of the standard exclusion process (a) and its phase diagram (b), where we emphasised in each phase a typical configuration of the lattice. The same process on a polymer, studied in this paper, is shown in panel (c), where $R$ is the end-to-end distance. Particles are represented by small discs, while  large discs represent the reaction volume of the entrance and exit sites (first and last site).}
	\label{fig::TASEP}
\end{figure}
\vspace{-2ex}
\subsection{Reminder of TASEP results}
In its simplest formulation the TASEP consists of a discrete lattice of $L$ sites where particles are injected from one end with rate $\alpha$, move from one site to the following one -if empty- with rate $p$, and are eventually removed from the last site with rate $\beta$, as illustrated in Fig.~\ref{fig::TASEP}(a). The system is usually studied by varying the dimensionless parameters $\bar\alpha := \alpha/p$ and $\bar\beta := \beta/p$, and the phase diagram of the system is known exactly~\cite{Blythe2007NonequilibriumGuide}. This is a rich model showing three different regimes (LD, low density; HD, high density; MC, maximal current), as well as first and second-order phase transitions. The LD-HD transition line shows a coexistence between the two phases, and it is often named SP (shock phase). The phase diagram is sketched in Fig.~\ref{fig::TASEP}(b), and each phase is characterised by a density of particles $\rho$ (average number of particles per site) and current $J$ (particles passing through a site per unit time). Mean-field approaches give the steady-state correct results, which we report in Table~\ref{eq::TASEP}, in the thermodynamic $L \rightarrow \infty$ limit.
\begin{table}[h!]
\centering
\begin{tabular}{c|c|c|c|c}
\hline
 Phase & Limits & Density & Current &  \\ \hline\hline
 LD & $\bar\alpha<\bar\beta, \bar\alpha< 1/2$ & $\rho = \bar\alpha$ & $J =  p\bar\alpha \left(1- \bar\alpha \right)$ &  \\ \hline
 HD & $\bar\alpha>\bar\beta, \bar\beta< 1/2$ & $\rho =  1-\bar\beta$ & $J =  p\bar\beta \left(1- \bar\beta \right)$ &  \\ \hline
 MC & $\bar\alpha,\bar\beta \geq 1/2$ & $\rho = 1/2$ & $J= p/4$ &  \\ \hline
\end{tabular}
\caption{Summary of the TASEP mean-field results.
\label{eq::TASEP} }
\end{table}

 We emphasise that, practically, exact densities and currents of lattices with a few tenths of sites can be reliably approximated with the mean-field expressions provided in Table~\ref{eq::TASEP}.

\section{Results} 

\subsection{Coupling TASEP on a polymer and a diffusive reservoir\label{subs:coupling}}
In the standard TASEP, the lattice is immersed in an infinite reservoir of particles for which the density determines the entry rate $\alpha$; the system can also be coupled to a finite reservoir of particles, and the effects of competition and depletion of a {\it homogeneous} reservoir without spatial extension has been tackled previously~\cite{Greulich2012}. Here instead we assume that the entry rate depends on the {\it local} concentration of particles $c$ in a reaction volume $V_a$ of radius $a$ around the first site, see Fig.~\ref{fig::TASEP}(c):
\begin{equation}
	\bar\alpha = \frac{\alpha_0}{p} \frac{1}{V_a}\int_{V_a} c(\mathbf{r}) d^3\mathbf{r} \,,
    \label{eq::int_alpha}
\end{equation}
where $c(\mathbf{r})$ is the concentration of particles in the reservoir at the position $\mathbf{r}$, and $\alpha_0$ plays the role of the reaction rate constant. In other words, particles are recruited in the lattice with a certain probability if they are at a distance smaller than $a$ from the entry site. 
For what concerns the lattice, in a worm-like chain polymer with persistence length $l_p$, the relation between its length $L$ and the mean-square of the end-to-end distance $R$ is given by~\cite{Rubinstein2003PolymerPhysics}
\begin{equation}
	R =   \left[2 l_p^2 \left(\frac{L}{l_p} -1+e^{-\frac{L}{l_p}}\right) \right]^\frac{1}{2} \sim\sqrt{2l_p L} \;,
    \label{eq::R}
\end{equation}
where we have approximated Eq.~(\ref{eq::R}) since $L \gg l_p$ in many practical cases. For instance, a typical mRNA of length $L=300$ codons has a persistence length of $\sim 1$ codon $\sim 1$ nm~\cite{Vanzi2005MechanicalComplexes}.  Although $R$ is a fluctuating variable, in what follows we limit ourselves to its mean-square value. The study of the role of fluctuations in this system goes beyond the scope of this paper. We briefly address this issue in Appendix~\ref{app:approx}. Other polymer models (like with self-avoiding monomer interactions) can also be implemented, as the reader can appreciate in Appendix~\ref{app:SAW}. This however does not change the conclusions of our work and the physics of the system.

For practical reasons we consider that the origin of the coordinate reference system coincides with the centre of the reaction volume surrounding the entry site.

In order to couple the transport process and the reservoir of particles we need to compute $c(\mathbf{r})$ inside $V_a$, which can be done by solving the diffusion equation with a sink centred at position $\mathbf{0}$ ($S_-$) and a source at position $\mathbf{R}$ ($S_+$):
  \begin{equation}
		D \nabla^2 c(\mathbf{r}) = S_+ (\mathbf{r}) - S_- (\mathbf{r}) \,,
		\label{eq::diff}
	\end{equation}
\noindent where $D$ is the diffusion coefficient of the particles in the reservoir. The sink and the source respectively describe the depletion, where particles are injected in the first site of the lattice, and their appearance around the last site where they abandon the unidimensional track. We exploit the steady-state condition, and considering for the sake of simplicity that the reaction volumes of sink and source are the same, we have $S_\pm (\mathbf{r}) := \pm J/V_a$. This connects the diffusion Eq.~(\ref{eq::diff}) to the TASEP currents in the three different phases (see Table~\ref{eq::TASEP}). We notice that the source $S_+$ term in Eq.~(\ref{eq::diff}) introduces a spatial feedback, which we also refer to as {\it particle recycling}, as particles leave the end site and, via diffusion, contribute to the local concentration inside the sink $S_-$.
 
The Poisson equation (\ref{eq::diff}) is the same holding in electrostatics to compute the potential $V(\mathbf{r})$ in a system with two spherical and homogeneous distributions of charges~\cite{Jackson1962, Eisler1969}. For an individual $S_\pm$, the concentration $c(\mathbf{x})$ at a distance $\mathbf{x}$ from the centre of each sphere can then be written as
\begin{equation}
	c(\mathbf{x}) = 
	\begin{cases}
		\pm\frac{J}{4\pi D |\mathbf{x}|} & \textrm{outside } S_\pm\\
		\pm\frac{J}{8\pi D a^3}(3a^2-x^2) & \textrm{inside } S_\pm \,.\\
\end{cases}
\label{eq::sol_source}
\end{equation}
By exploiting the linearity of the diffusion equation and fixing the density far away from the lattice to be $c_\infty$, we construct the expression of $c(\mathbf{r})$ used to compute the entry rate. Solving the integral in Eq.~(\ref{eq::int_alpha}), we obtain the expression for the injection $\bar\alpha$ as a function of the current $J$ and the distance $R$ between the entry and exit sites:
\begin{equation}
	\bar\alpha = \bar\alpha_\infty +  \frac{J}{p} \Gamma\,,
\label{eq::alpha}
\end{equation}
where $\bar\alpha_\infty := \alpha_0 c_\infty/p$ corresponds to the injection parameter usually considered in standard TASEP-based models (i.e. without particle recycling and reservoir depletion), and 
\begin{equation}
    \Gamma := 
    \begin{cases}
        \displaystyle\frac{\alpha_0}{4\pi D a}  \left(\frac{1}{2d} - \frac{6}{5} \right)\qquad &\textrm{for  } d\geq 1 \\
        \displaystyle\frac{\alpha_0}{4\pi D a}  d^2 \left[\frac{3}{2}d - \frac{1}{5}d^3 -2 \right] \qquad &\textrm{for  } d<1 \,,
    \end{cases}
    \label{eq::Gammas}
\end{equation}
where $d := R/2a$. For the derivation of the previous Equations we refer to Appendix~\ref{app:comp}. We recover the standard TASEP when $\Gamma = 0$, i.e. when $D\rightarrow \infty$ and we can neglect the spatial inhomogeneities in the reservoir.

Coupling Eq.~(\ref{eq::alpha}) to the particle current in the LD, HD and MC phases shows how these different regimes affect the spatial feedback and thus the injection $\bar\alpha$. Equation~(\ref{eq::alpha}) will therefore take different forms according to the phase of the TASEP (see Table~\ref{eq::TASEP}):
\begin{equation}
	\bar\alpha = 
	\begin{cases}
	\bar\alpha_\infty + \bar\alpha(1-\bar\alpha) \, \Gamma &  \textrm{(LD)}\\
	\bar\alpha_\infty + \bar\beta(1-\bar\beta) \, \Gamma &  \textrm{(HD)}\\
	\bar\alpha_\infty + \Gamma/4 &  \textrm{(MC)}\;.\\
	\end{cases}
    \label{eq::alpha_phases}
\end{equation}
Only in the LD phase we need to solve a second order equation to find $\bar\alpha$ and obtain
\begin{equation}
\bar\alpha_{LD} = \frac{(\Gamma - 1)\pm\sqrt{(1-\Gamma)^2+4\bar\alpha_{\infty}\Gamma}}{2\Gamma}      \,.
\label{eq::alpha_LD}
\end{equation}
We recall that the solution $\bar\alpha_{LD}$ is relevant only if $\bar\alpha<\bar\beta$ and $\bar\alpha<1/2$ (otherwise the system is in HD or MC); we always find only one physical solution $\bar\alpha_{LD}$.

\begin{figure}[t]
	\centering
	\includegraphics[width=.99\columnwidth]{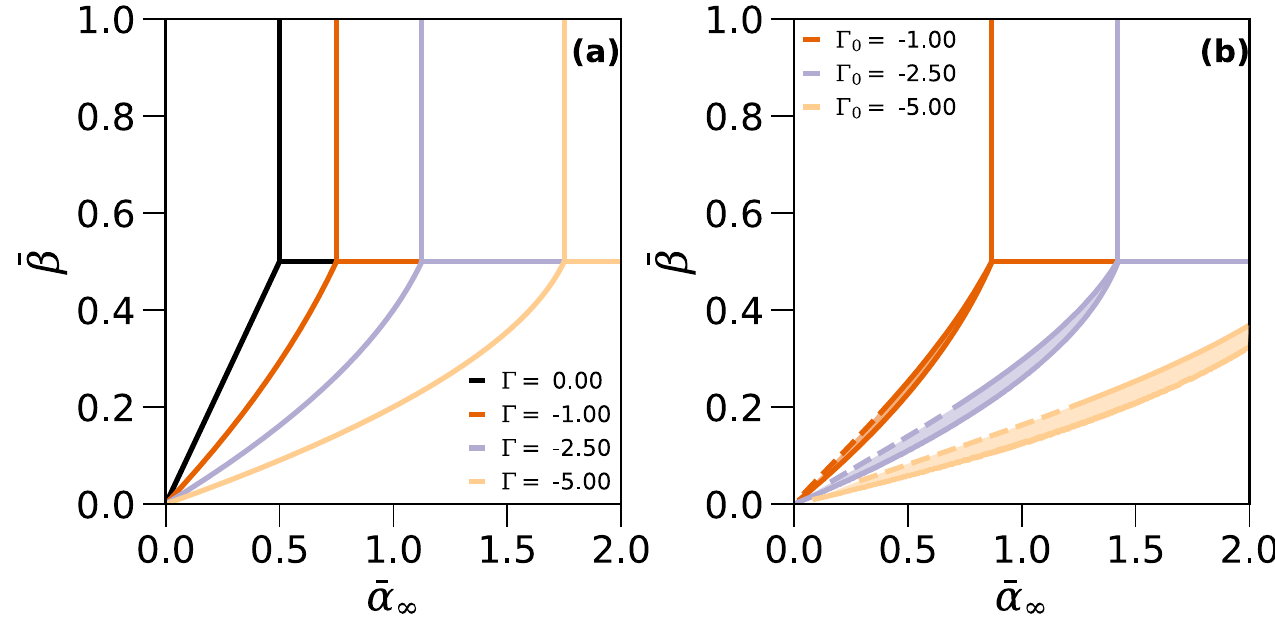}  
	\caption{ (Color online) (a) Phase diagram of the TASEP coupled with the diffusive reservoir in the $\{\bar\alpha_\infty, \bar\beta\}$ plane for different values of $\Gamma$, and (b) phase diagram of the TASEP {\it on a flexible polymer} for different values of $\Gamma_0$, fixing $a=1.5$, $\ell=1$, $l_p =0.1$ and $L=25$ (giving $R_0\simeq 2.24$). All lengths are expressed in lattice-site units. We plug Eqs.(\ref{eq::Gamma_pol}) in (\ref{eq::alpha_phases}) to compute the transition lines; when the lattice is compact $d<1$ we use dashed lines and a full line otherwise. The shaded areas in between the LD$\rightarrow$SP and HD$\rightarrow$SP transition lines highlight the extended coexistence region.}
	\label{fig::phase_diagram}
\end{figure}

The phase boundaries given in Table~\ref{eq::TASEP} can be rewritten in terms of the new parameters and in Fig.~\ref{fig::phase_diagram}(a) we show the phase diagram of the system in the $\{\bar\alpha_\infty, \bar\beta\}$ plane for different values of $\Gamma$.
As expected from Eq.~(\ref{eq::alpha}), if $\Gamma = 0$ we recover the phase diagram of the TASEP (black lines) with $\bar\alpha_\infty$ playing the role of the entry rate of the standard TASEP. This situation is also met when the reaction volumes of the entrance and exit sites match, i.e. for a fully circularised lattice suppressing the depletion of the reservoir~\cite{Fernandes2018InPreparation}.
The dimensionless parameter $\Gamma$ is otherwise always negative, and it weights the interplay between particle recycling and depletion around the entrance site of the lattice, which is proportional to $J$.
When $\Gamma < 0$ the MC regime is reached for increasingly larger values of $\bar\alpha_\infty$.\\

\subsection{Coupling polymer conformation and transport} 
In the previous sections we assumed that the transport process does not interact with the lattice conformation. Here we impose that the particle density stiffens the polymer and hence controls the global configuration of the lattice. In turn this will have a repercussion on the particle recycling that depends on $R$. 

If the persistence length of the lattice $l_p$ is larger than the particle's footprint $\ell$, the transport process does not affect the end-to-end distance. Instead, if $\ell/2 \geq l_p$, the presence of a particle on the lattice flattens the region of the lattice corresponding to its footprint; as a consequence, the particle density $\rho$ influences the end-to-end distance, changing the features of particle recycling. To compute the effective persistence length $l_\textrm{eff}$ of a lattice covered with particles we consider a freely-jointed chain composed by fragments of polymers with two different Kuhn lengths: $\ell$ that corresponds to a particle footprint, and $2l_p$ for the empty lattice. In the limit $L$ much greater than the Kuhn length one recovers the results for the end-to-end distance Eq.~(\ref{eq::R}) of the worm-like chain model mentioned in Section \ref{subs:coupling}, with a Kuhn segment equal to twice the persistence length. Knowing that the portion of the lattice occupied by particles is $\rho \ell$, and that the portion of the empty lattice is ($1-\rho \ell$), we obtain $l_\textrm{eff} = \rho \frac{\ell^2}{2} + (1-\rho\ell)l_p$, assuming orientation independence between particle-occupied and free-chain fragments. This relation, which, for the sake of clarity we derive in Appendix~\ref{app:leff}, is valid only when $\ell/2 \geq l_p$, and otherwise $l_\textrm{eff} = l_p$.
We can then redefine 
\begin{equation}
R := \sqrt{2l_\textrm{eff} L} = R_0 F_\rho \,,
\label{eq::R_polymer}
\end{equation}
where 
\begin{eqnarray}
R_0 & := & \sqrt{2l_p L}  \\ 
F_\rho & := & \left[1 + \rho\ell\left( \frac{\ell}{2\,l_p}-1\right) \right]^\frac{1}{2}\,,\quad  \ell/2 \geq l_p\,.
\label{eq::R_pol}
\end{eqnarray}
Here $R_0$ is the mean end-to-end distance of an empty polymer and $F_\rho = R/R_0\geq 1$ is a measure of how much the polymer is flattened by the particle occupancy. We use either $R$ or $F_\rho$ as proxies for the polymer conformation. The parameter $\Gamma$ then reads
\begin{equation}
    \Gamma = 
    \begin{cases}
        \Gamma_0 + \displaystyle\frac{\alpha_0}{4\pi D} \frac{1}{R_0}\left( \frac{1-F_\rho}{F_\rho}\right) & d \geq 1 \\
        \left[ \Gamma_0 + \frac{\alpha_0 d_0^3}{4\pi D a}\left[ \frac{3}{2}(F_\rho-1) - \frac{1}{5}d_0^2 (F_\rho^3-1)\right] \right] F_\rho^2 & d < 1 \,,
    \end{cases}
    \label{eq::Gamma_pol}
\end{equation}
where $\Gamma_0$ is obtained by calculating the parameter $\Gamma$ from Eqs.~(\ref{eq::Gammas}) by setting $R=R_0$ and $d_0:= R_0/2a$.
We emphasise that we recover the results of the previous section when $\ell/2 \leq l_p$ and therefore the transport process and polymer properties are decoupled.
\begin{figure}[tb] 
	\centering
	\includegraphics[width=.99\columnwidth]{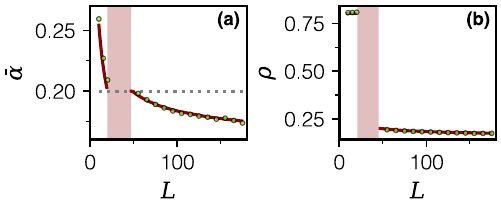}  
	\caption{ Entry parameter $\bar\alpha$ (a) and density $\rho$ (b) as a function of the lattice length $L$, with $\bar\alpha_\infty = 0.5$, $a=1.5$, $\ell=1$, $l_p =0.1$ and  $\alpha_0/(4 \pi D a) = 2.25$, $\bar\beta = 0.2$ (indicated by the dashed line). When $\bar\alpha > \bar\beta$ ($\bar\alpha < \bar\beta$) the system is in HD (LD); other values of $L$ (shaded area) correspond to the extended coexistence region of the phase diagram - Fig.~\ref{fig::phase_diagram}(b).}
	\label{fig::length}
\end{figure}
We can still use Eqs. (\ref{eq::alpha_phases}) to compute the entry parameter in the different phases, although, this time, $\Gamma$ is a function of $\rho$, which also depends on the phase as given in Table~\ref{eq::TASEP}. For instance, to find the value of $\bar\alpha$ in LD, now we need to solve $\bar\alpha =  \bar\alpha_\infty + \bar\alpha(1-\bar\alpha)\Gamma$ with $\Gamma$ from Eq.~(\ref{eq::Gamma_pol}) computed with $\rho = \bar\alpha$ (values for HD are calculated using the respective values of density and current, see Table~\ref{eq::TASEP}).
By solving those equations it is possible to determine the phase boundaries of the three TASEP phases, now considering the feedback that the polymer conformation dictates on the exclusion process. We show this phase diagram in Fig.~\ref{fig::phase_diagram}(b) for different values of $\Gamma_0$ and for parameters that could represent ribosome translating an mRNA (see Appendix~\ref{app:mRNA}).

This phase diagram shows a remarkable new feature, i.e. the presence of an extended LD-HD coexistence region, emerging by the competition between particle recycling and stiffening of the polymer. When coupling transport and polymer conformation, entering the SP from LD or from HD generates two different transition lines since $\Gamma$ in Eqs.(\ref{eq::alpha_phases}) depends on the density as in Eqs.(\ref{eq::Gamma_pol}). In the extended SP we expect on average $\bar\alpha = \bar\beta$, and hence $\Gamma$ will be fixed in order to meet this condition. 

Figure~\ref{fig::length} illustrates how transport is affected by the length $L$ of the lattice, via the coupling between particle recycling and the substrate's conformation. We show how $\bar\alpha$ and $\rho$ change with the system-size and compare the theoretical prediction (lines) to the outcome of numerical simulations running a TASEP with a particle injection rule given by Eq.(\ref{eq::alpha}). This length-dependence is absent in the standard TASEP, where steady-state quantities are independent of the system size. Details of the simulations are summarised in Appendix~\ref{app:sim}.

\subsection{Polymer conformation as a proxy for transport regimes} 
\begin{figure}[ht] 
	\centering
	\includegraphics[width=1\columnwidth]{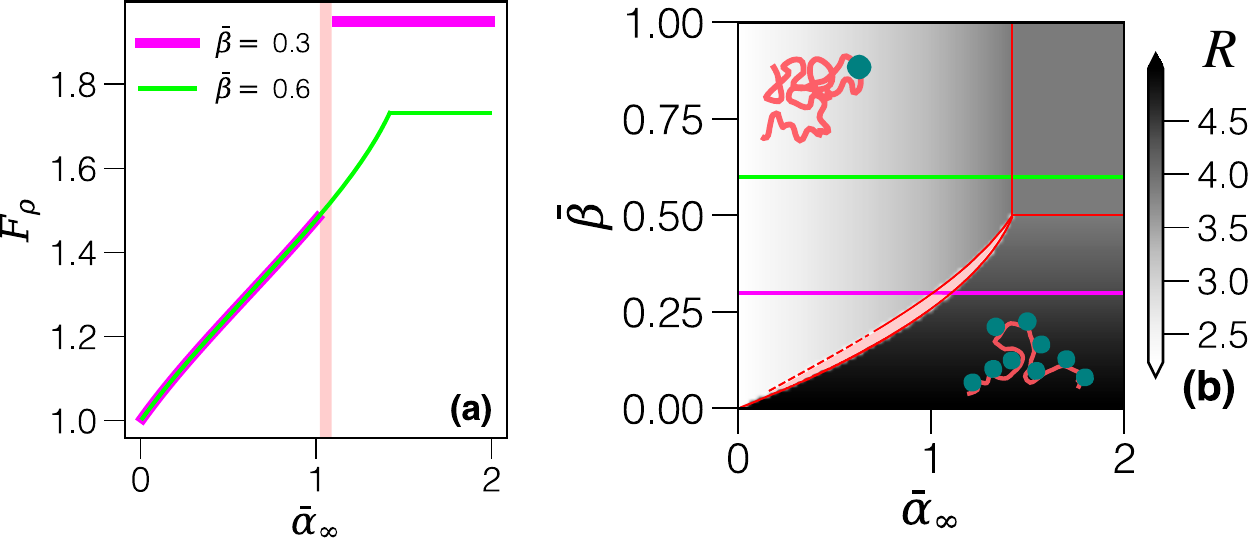}  
	\caption{(Color online) Transport effects on polymer conformation. Panel (a) shows $F_\rho=R/R_0$ as a function of $\bar\alpha_\infty$, with the shaded region indicating the extended SP. In (b), a color map of the end-to-end distance $R$ (in lattice-site units) in the $\{\bar\alpha_\infty, \bar\beta\}$ plane. The horizontal lines correspond to $\bar\beta=0.3$ (magenta, dark gray) and $\bar\beta=0.6$ (green, light gray), for which $F_\rho$ is shown in panel (a) maintaining the same color code. In both panels $\Gamma_0 = -2.5$ and the other parameters correspond to the ones used in Fig.~\ref{fig::phase_diagram}(b). Red (dark thin) lines correspond to the theoretical phase diagram. In the upper left and bottom right corners of panel (b) we sketch cartoons of the polymer configurations with particles (small discs) on it corresponding to those regions of the diagram: stretched polymers (large $R$) are found in the TASEP HD phase, while more compact lattices are expected in the LD (compare phases of the standard TASEP phase diagram shown in Fig.~\ref{fig::TASEP}).
	}
	\label{fig::colormap}
\end{figure}

The density of particles $\rho$ thus impacts the typical polymer conformation as given in Eq.~(\ref{eq::R_polymer}). As shown in Fig.~\ref{fig::colormap}(a), obtained by numerically solving Eq.(\ref{eq::R_pol}), with increasing $\bar\alpha_\infty$, we observe that the polymer conformation undergoes a transition from a compact to a more flattened shape ($F_\rho>1$) driven by the accumulation of particles on it.  
We emphasise that $\bar\alpha_\infty$ is proportional to the particle concentration, meaning that features of the polymer conformation will vary by changing the particle concentration $c_\infty$.

In Fig.~\ref{fig::colormap}(b) we show how regimes of the polymer conformation coincide to the different phases of the exclusion process.
We can appreciate that $R$ follows the different dynamical TASEP phases: when the system is in LD, the polymer is in its compactest shape, while the ends get separated in the MC and the polymer becomes more and more stretched deep in the HD phase, as sketched in Fig.~\ref{fig::colormap}(b).

\section{Discussion}
We have developed a novel approach to study the interdependence between transport on a unidimensional substrate (the exclusion process) and the three-dimensional conformation of the lattice on which the particles move. 
We propose a coupling between driven transport and polymer dynamics that influences the three-dimensional conformation of the polymer, and thus particle recycling. Our model then couples active transport in 1D and passive transport in 3D, expanding previous investigations on purely 1D systems~\cite{Hinsch2006}.

In our approach the physical properties of the uni-dimensional lattice thus play a fundamental role.
We show that, in this perspective, features of the lattice such as its length or persistence length cannot be overlooked --as usually done in coarse grained modelling-- to provide a complete description of the transport process.  The conformational state of the polymeric lattice becomes informative of the properties of the transport occurring on it. Thus, the lattice conformation could be exploited to estimate regimes of transport.  
For instance, there is evidence that typical structures of polysomes (mRNAs with active ribosome translating) depend on the transport process and in particular on the ribosome recruitment~\cite{Viero2015ThreePolysomes.} (this effect can be experimentally achieved by antibiotics, which effectively decrease the amount of ribosomes $c_\infty$~\cite{Greulich2015Growth-dependentAntibiotics}). Furthermore, coupling between particle transport and polymer conformation might also be important in DNA transcription factories, where the local concentration of polymerases and their recruitment are relevant.

The role of mRNA circularisation in determining gene expression is still largely unknown. The full circularisation of eukaryotic transcripts is assisted by molecular partners promoting the interaction between their ends; when this interaction is disrupted the translation efficiency strongly decreases~\cite{Wesselhoeft2018}. This is consistent with our model, which predicts optimal ribosome recycling with full circularisation ($\Gamma = 0$).  The formation of the circularised state, however, competes with the stiffening of the polymer induced by high density translation, which reinforces the importance of considering the process on a flexible substrate.

Finally, by considering particle recycling, and the interplay between transport and polymer conformation, we show an extended coexistence region in the phase diagram, a feature that is not present in the standard TASEP. Due to its similarity with the phase diagrams shown in~\cite{Greulich2012, Brackley2012MultipleCarriers}, we hypothesize that in this regime, in order to maintain $\bar\alpha = \bar\beta$, a pinned domain wall might emerge in the lattice to adapt the density, and hence allow $\Gamma$ to satisfy the constraint.

Dynamical effects that could be present in the LD-HD coexistence region may be addressed by Molecular Dynamics simulations. Future works might also explore extensions to inhomogeneous TASEP~\cite{Szavits-Nossan2018DecipheringRate,*Szavits-Nossan2018PowerProcess}, finite resources~\cite{Greulich2012}, TASEP with extended particles~\cite{Shaw2003TotallySynthesis,*Erdmann-Pham2018HydrodynamicSolution} or particles that modify local curvature differently from the flattening that we considered.\\

{\it \noindent Acknowledgements.}  LDF acknowledges the funding provided by the S\~{a}o Paulo Research Foundation (FAPESP - grant 2015/26989-4) and LC  the CNRS for having granted him a ``demi-d\'el\'egation''. The authors thank Philip Greulich and Norbert Kern for their useful comments on the manuscript. We would like to dedicate this work to the memory of Bruno Bassetti.

\appendix
\section{Summary of the approximations used \label{app:approx}}
In this section we summarise the approximations behind the modelling framework we have developed. For the sake of clarity we explain the assumptions made in the two sections of the main text were we developed the theory.

\subsection{Coupling TASEP on a polymer and a diffusive reservoir}
\begin{figure}[htbp]
\centering
\includegraphics[width = 0.9\columnwidth]{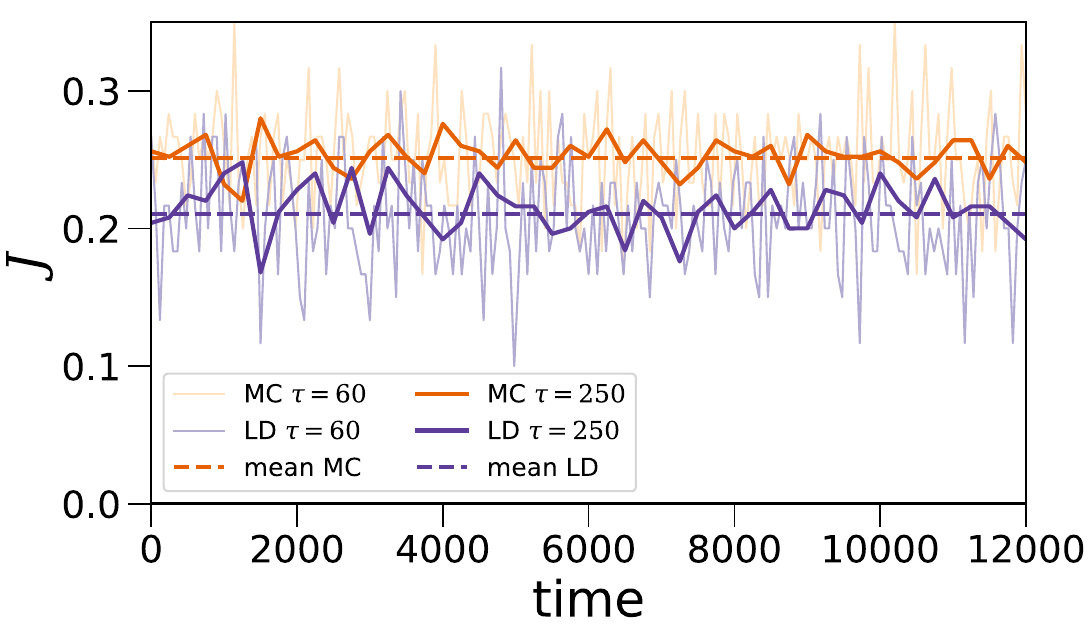}
\caption{(Color online) Current $J(t)$ for a TASEP in LD ($\alpha =0.3$  s$^{-1}$, $\beta = 0.8$ s$^{-1}$) and in MC ($\alpha = 0.7$  s$^{-1}$, $\beta = 0.7$ s$^{-1}$). For both cases $p=1$ s$^{-1}$ and $L=25$ (codons). Currents are computed every time interval $\tau = 60\ \mbox{s}$ or $\tau = 250\ \mbox{s}$. 
}
\label{current}
\end{figure}

In the corresponding section of the main text we have assumed the validity of mean field for the concentration field $c(\mathbf{r})$. This holds when the sources $S\pm$ do not show large fluctuations in time and can be considered to produce/absorb a constant number of particles per unit time. In fact, this can be verified with simulations of the TASEP. Since $J(t)$ is obtained as an average over time windows, fluctuations are, however, strongly depend on the size $\tau$ of the time window chosen. This can be qualitatively examined in Fig. \ref{current}, for $\tau = 60\ \mbox{s}$ and $250\ \mbox{s}$. Fluctuations, of course, increase for decreasing values of $\tau$ and in the limit $\tau\rightarrow 0$ fluctuations are maximal, when measurements of the current would give a straight line at $0$ and spikes at $1$ when particles exit.

\begin{figure}[ht]
\begin{center}
\includegraphics[width = 0.9\columnwidth]{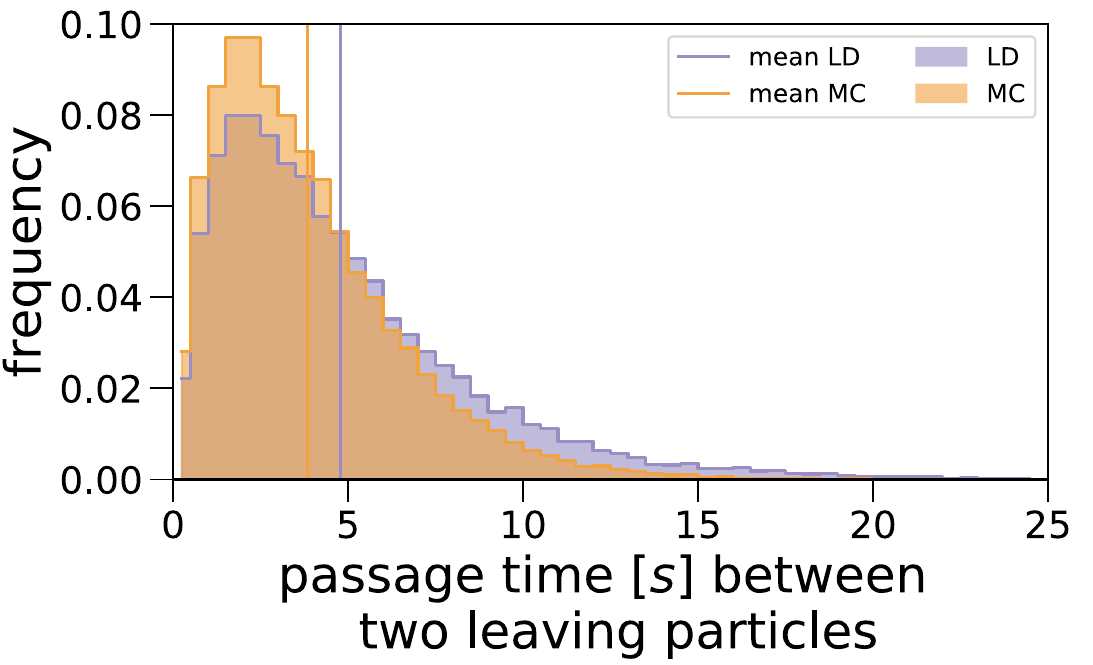}
\caption{(Color online) Histograms of passage times (in seconds) between two consecutive particles leaving the lattice (same parameters of Fig. \ref{current}). The vertical lines represent the measured $1/J$ values, giving the timescale of the current.}
\label{hist}
\end{center}
\end{figure}

To avoid the arbitrariness of the value of $\tau$ used to compute the observables, we use another discriminant to determine the validity of our approximation. We consider the {\it passage time} between two leaving particles. The idea is that if two particles exit too close, the reservoir and the diffusion in the reservoir will not be able to absorb strong fluctuations of the current, and hence $\delta c \sim c$ ($\delta c$ being the fluctuations in the density). A lower bound for the passage time should be fixed by the diffusion timescale $\tau_D$, set as the average time for a particle to diffuse a distance equivalent to the average end-to-end distance of the transcript. Thus, if couples of particles have average passage times smaller than $\tau_D$, fluctuations in the concentration field are likely to be relevant and the mean-field approximation is less likely to hold. Fig. \ref{hist} shows the histograms of passage times for two different phases of the TASEP. Average passage times in both phases are around $4-5$ s. On the other hand, we estimate $\tau_D$ for translation of being of the order $10^{-3}$ s ($\tau_D=R^2/6D=2Ll_p/6D$, and the diffusion coefficient of ribosomes $D=0.04\ \mbox{$\mu$m}^2/\mbox{s}$~\cite{Bakshi2012SuperresolutionCells}). For transcription, $\tau_D$ might be obtained in a similar fashion.

Thus, with an average passage time orders of magnitude larger than the diffusion timescale, we expect that fluctuations in the concentration might not play a significant role and that, therefore, our approximation is likely to hold.\\

We have also implicitly assumed that the timescale of diffusion of the polymer position and of its ends is smaller than particle diffusion, otherwise the reservoir will be closer to a well-mixed assumption: coupling polymer and transport would still be meaningful but recycling could be neglected.\\

To avoid misunderstanding, we remind that diffusion is modelled as continuous process in 3D, and transport is described on a 1D lattice following standard TASEP rules; the two systems are coupled as explained above and in the main text, and the 1D lattice does not interfere with diffusion.

\subsection{Coupling polymer conformation and transport}
In that section of the main text we couple transport and polymer conformation, and in particular we study how particle density can affect the end-to-end distance of the lattice. 

As we mentioned in the Introduction section, we consider that locally the polymer moves much faster than the particle hopping, and we can then decouple the dynamics. Substantially the polymer equilibrates faster after each particle step. In this work we have focused on this regime. Other regimes, although might hide interesting dynamical effects, are out of the scope of this work.\\

This approximation also allows us to neglect fluctuations in the end-to-end distance $R$, that could be taken into account in extensions of the model. For instance one could modify Eq.(5) of the main text and compute $\langle J \Gamma \rangle$, where the brackets represent the average over time. In our approximation, thanks to the timescale separation we have implicitly assumed $\langle J \Gamma \rangle \sim \langle J \rangle \langle \Gamma \rangle = J \, \Gamma$. This means that we can decouple TASEP and polymer dynamics, and that the average end-to-end distance $R$, Eq.(2), is representative of the distance between entry and exit sites.\\

\begin{figure}[htbp]
\begin{center}
\includegraphics[width = 0.9\columnwidth]{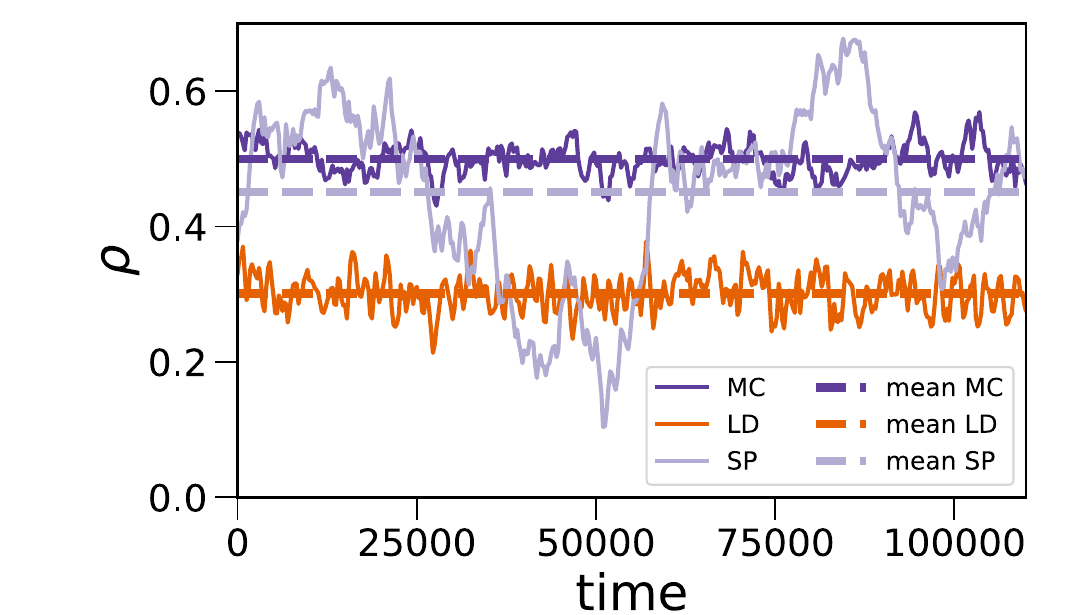}
\caption{(Color online) Density of particles as a function of time for LD and MC phases and the SP regime, with $\tau =250\ \mbox{s}$. For the SP regime $\alpha=\beta=0.1$ s$^{-1}$, and the other parameters are the same as the ones used in Fig.~\ref{current}.\label{density}}
\end{center}
\end{figure}

We want to stress that, with this assumption, we can relate $\rho$ and $R$. That is also possible thanks to the steady-state assumption. This implies that fluctuations are small compared to the average, and then the approximations used in this work generally hold. However, we bring to the reader's attention that at the transition between LD and HD phases (often named Shock Phase, SP), the average density fluctuates in time~\cite{Blythe2007NonequilibriumGuide}: a domain wall in the density profile is present and links the LD and HD density that coexist in the lattice when this first order transition occurs. Interestingly, in the standard TASEP the position of the domain wall makes a random walk on the lattice, which should generate large fluctuations in the lattice conformation. In Fig. \ref{density} we include plots of the simulated $\rho(t)$ in the LD, MC phases and SP regime, showing that only in the SP large fluctuations can constitute important dynamical effects. Hence, the identification of the transition lines might become inaccurate and the SP region might become even larger if fluctuations dominate. Instead, if in the extended region a pinned domain wall is present, fluctuations will decrease and the theoretical location of the phase boundaries will be precise. However, a deeper investigation of this region and of fluctuations is out of the scope of this paper, and they could be addressed with explicit simulations of polymer dynamics.

Moreover, we consider the case in which the particle hopping is not influenced by the polymer state. For instance we do not consider local secondary structures of the lattice, assuming that the particles can efficiently unfold them or that folding competes with the particle flow~\cite{Turci2013TransportDefects} (as in the case of ribosomes moving on the mRNA). 

\vspace{-3ex}
\section{Derivation of initiation rate\label{app:comp}}

\vspace{-2ex}
\subsection{Non-intersecting reaction spheres ($R > 2a$)}

As stated in the main text, the entry rate $\alpha$ is given by the integral of the concentration of particles (considered as point particles) over the reaction volume $V_a$, centered at the initiation site:
\begin{equation}
	\alpha = \alpha_0 \frac{1}{V_a} \int_{V_a} c(\mathbf{r}) d^3\mathbf{r} \,.
\label{eq::initiation}
\end{equation}

By the principle of superposition, taking the concentration terms of equation (4) of the main text, the concentration inside the reaction volume $S_{-}$ will be given by:
\begin{equation}
 c(\mathbf{r})  = c_\infty  + \displaystyle\frac{J}{4\pi D a}\left( \frac{a}{|\mathbf{r} - \mathbf{R}|} + \frac{|\mathbf{r}|^2}{2a^2} -\frac{3}{2} \right)
 \label{con01}
\end{equation}

Eq.~(\ref{con01}) can be readily integrated to calculate $\alpha$ following Eq.~(\ref{eq::initiation}). With the reaction volume $S_{-}$ centered at the origin, we have: 
\begin{widetext}
\begin{align}
	\alpha &= \alpha_0 \frac{1}{V_a} \int_{S_{-}} \left[ c_\infty  + \displaystyle\frac{J}{4\pi D a}\left( \frac{a}{|\mathbf{r} - \mathbf{R}|} + \frac{|\mathbf{r}|^2}{2a^2} -\frac{3}{2} \right)\right] d^3\mathbf{r} \notag \\
    \alpha &= \alpha_0 \frac{1}{V_a} \int_{0}^{2\pi}\int_{0}^{\pi}\int_{0}^{a} \left[ c_\infty  + \displaystyle\frac{J}{4\pi D a}\left( \frac{a}{\sqrt{r^2+R^2-2rR\mbox{cos}\theta}} + \frac{r^2}{2a^2} -\frac{3}{2} \right)\right] r^2\sin{\theta} dr d\theta d\phi \,.
    \label{int_alf}
\end{align}
\end{widetext}

\noindent The solution of the integral \ref{int_alf} gives:
\begin{equation}
\alpha = \alpha_0\left[c_\infty + \frac{J}{4\pi D}\left(\frac{1}{R} - \frac{6}{5a}\right) \right] \,.
\end{equation}

\subsection{Intersecting reaction spheres ($R < 2a$)}

When $R<2a$, there is an intersection between the source (reaction volume centered at the termination site, $S_{+}$) and the sink (centered at the initiation site, $S_{-}$). The main issue to solve is how to calculate the integral given by Eq.~(\ref{eq::initiation}).

By the superposition of the solutions of the Laplace equation, with the origin at the centre of the region $S_{-}$, we have :
\begin{equation}
	c(\mathbf{r}) = 
	\begin{cases}
		c_\textrm{out}(\mathbf{r})  & = c_\infty  + \displaystyle\frac{J}{4\pi D a}\left( \frac{a}{|\mathbf{r} - \mathbf{R}|} + \frac{|\mathbf{r}|^2}{2a^2} -\frac{3}{2} \right) \\ &\textrm{in } S_- \textrm{ outside the intersection}\\
		c_\textrm{int}(\mathbf{r})  &=c_\infty  - \displaystyle\frac{J}{8\pi D a^3}\left[ R^2- 2(\mathbf{r} \cdot \mathbf{R})  \right] \\ & \textrm{in } S_- \textrm{ inside the intersection.}
	\end{cases}
\label{eq::sol_diff}
\end{equation}

\noindent To obtain the last line of Eq.(\ref{eq::sol_diff}) we added the terms of $S_-$ and $S_+$ inside the respective reaction volumes:
\begin{eqnarray}
	c_\textrm{int}(\mathbf{r}) &= & c_\infty  - \frac{J}{8\pi D a^3}\left( 3a^2- r^2  \right) + \frac{J}{8\pi D a^3}\left( 3a^2- |\mathbf{r} - \mathbf{R}|^2 \right) \notag \\
	& = & c_\infty  - \frac{J}{8\pi D a^3}\left[ R^2 - 2(\mathbf{r} \cdot \mathbf{R}) \right] \,,
\end{eqnarray}

\noindent where $-\frac{J}{8\pi D a^3}\left( 3a^2- r^2  \right)$ and $\frac{J}{8\pi D a^3}\left( 3a^2- |\mathbf{r} - \mathbf{R}|^2 \right)$ are the contributions by the sink and the source, respectively, to the concentration in the intersection region.

The integral in Eq.(\ref{eq::initiation}) can then be computed as it follows:
\begin{eqnarray}
	\alpha =& \alpha_0 \frac{1}{V_a} \int_{V_a} c(\mathbf{r}) d^3\mathbf{r} \nonumber  \\  \nonumber
	= & \alpha_0 \frac{1}{V_a} \left( \int_{V_a} c_\textrm{out}(\mathbf{r}) d^3\mathbf{r} \right.\\ 
	& + \left. \int_{V_{\textrm{int}}} (-c_\textrm{out}(\mathbf{r}) + c_\textrm{int}(\mathbf{r})) d^3\mathbf{r} \right)\, \,.
\label{eq::initiation2}
\end{eqnarray}

Note that, by symmetry, the integral over the intersection volume of the contributions of source and sink cancel out. Thus,
\begin{equation*}
    \int_{V_{\textrm{int}}} c_\textrm{int}(\mathbf{r}) d^3\mathbf{r} =  \int_{V_{\textrm{int}}} c_\infty d^3\mathbf{r} \,.
\end{equation*}

Equation~(\ref{eq::initiation2}) will then be given by:
\begin{widetext}
\begin{eqnarray}
	\alpha & = & \alpha_0 \frac{1}{V_a} \left( \int_{V_a} c_\textrm{out}(\mathbf{r}) d^3\mathbf{r} + \int_{V_{\textrm{int}}} (-c_\textrm{out}(\mathbf{r}) + c_\textrm{int}(\mathbf{r})) d^3\mathbf{r} \right) \notag \\
	& = & \alpha_0 \frac{1}{V_a} \left( \int_{V_a} c_\textrm{out}(\mathbf{r}) d^3\mathbf{r} - \int_{V_{\textrm{int}}} (c_\textrm{out}(\mathbf{r}) - c_\infty) d^3\mathbf{r} \right) \notag \\
    & = & \alpha_0 \frac{1}{V_a} \left( \int_{V_a} \left(c_\textrm{out}(\mathbf{r})-c_\infty +c_\infty\right) d^3\mathbf{r} - \int_{V_{\textrm{int}}} (c_\textrm{out}(\mathbf{r}) - c_\infty) d^3\mathbf{r} \right) \notag \\
    & = & \alpha_0 \frac{1}{V_a} \left( \int_{V_a} \left(c_\textrm{out}(\mathbf{r})-c_\infty\right) d^3\mathbf{r} + \int_{V_a} c_\infty d^3\mathbf{r} - \int_{V_{\textrm{int}}} (c_\textrm{out}(\mathbf{r}) - c_\infty) d^3\mathbf{r} \right) \notag \\
	& = & \alpha_0 \frac{1}{V_a} \left( \int_{V_a} c_\infty d^3\mathbf{r} + \int_{V_a-V_{\textrm{int}}} (c_\textrm{out}(\mathbf{r}) - c_\infty) d^3\mathbf{r} \right) \notag \\
	& = & \alpha_0 c_\infty + \alpha_0 \frac{1}{V_a} \left(\int_{V_a-V_{\textrm{int}}} (c_\textrm{out}(\mathbf{r}) - c_\infty) d^3\mathbf{r} \right) \, ,
\label{eq:alpha_gen}
\end{eqnarray}
\end{widetext}

\noindent where $V_a-V_{\textrm{int}}$ corresponds to the region inside $S_{-}$ excluding the intersection region.

\begin{figure}[!ht]
	\centering
	\includegraphics[width=.8\columnwidth]{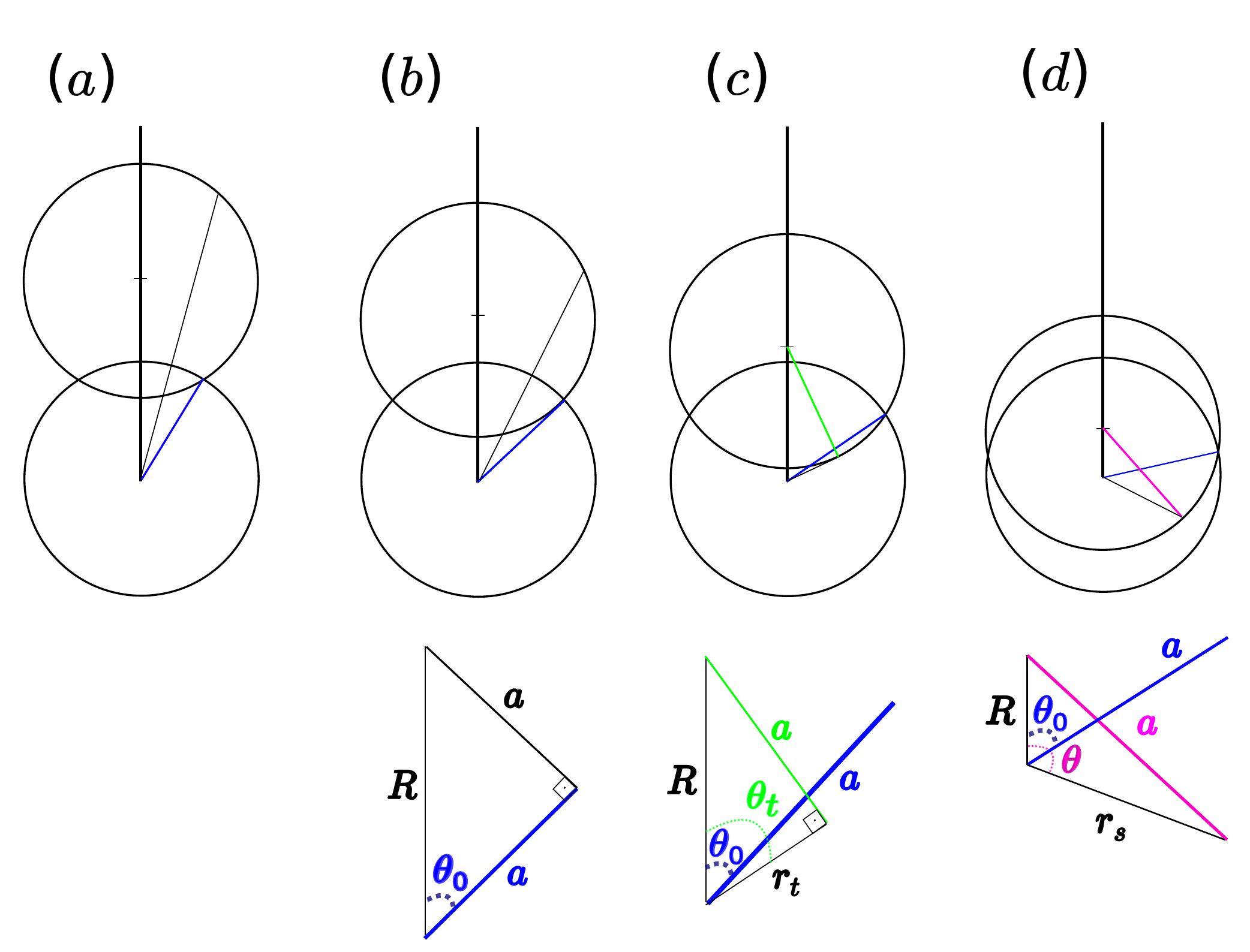}  
	\caption{Sketches of two intersecting spheres with varying distance $R$ between the centres.}
	\label{fig::sph_int}
\end{figure}

Aligning the $z$-axis in the direction of the vector $\mathbf{R}$ (which gives the position of the centre of the source ($S_{+}$), the limits of integration for Eq.~(\ref{eq:alpha_gen}) present three cases of interest: $a\sqrt{2} < R <2a$ (Fig.~\ref{fig::sph_int}(a) and (b)), $a < R < a\sqrt{2}$ (Fig.~\ref{fig::sph_int}(c)) and $R<a$ (Fig.~\ref{fig::sph_int}(d)). Integration in spherical coordinates for each of these cases is discussed below.

\subsubsection{Case $a\sqrt{2} < R < 2a$}

From the equation of the sphere of the source we have:
\begin{align}
 &x^2+y^2+(z-R)^2 = a^2 \notag \\
 &x^2+y^2+z^2-2zR+R^2 = a^2 \notag
\end{align}

\noindent but since $r^2=x^2+y^2+z^2$ and $z=r\cos{\theta}$, we have:
\begin{align}
 &r^2-(2R\cos{\theta})r+(R^2-a^2) = 0 \notag \\
 &r_{-} = R\cos{\theta}-\sqrt{R^2\cos^2{\theta}-(R^2-a^2)},
\end{align}

\noindent (note that since $R>a$ for each value of $\theta$ there are two values: $r_{+}$ and $r_{-}$. The ``$+$''-solution corresponds to the larger value of $r$).

The polar angle $\theta_0$ corresponds to the maximum aperture in the intersection region (Fig.~\ref{fig::sph_int}(a) and (b)) and is given by:
\begin{equation}
 \cos{\theta_0}=\frac{R}{2a} \, ,
\end{equation}

\noindent since the line passing through the centre of the intersection crosses the $z$-axis at $z=R/2$. For the region of the sphere outside the intersection we then have:
\begin{widetext}
\begin{align}
 \int_{V_a-V_{\textrm{int}}} (c_\textrm{out}(\mathbf{r}) - c_\infty) d^3\mathbf{r} =& \displaystyle\frac{J}{4\pi D a} \left\{\int_0^{2\pi}\int_0^{\theta_0}\int_0^{r_{-}} \left( \frac{a}{|\mathbf{r} - \mathbf{R}|} + \frac{|\mathbf{r}|^2}{2a^2} -\frac{3}{2} \right) r^2\sin{\theta} dr d\theta d\phi \ + \right. \notag \\
 & \hspace{2cm}+ \left. \int_0^{2\pi}\int_{\theta_0}^{\pi}\int_0^{a} \left( \frac{a}{|\mathbf{r} - \mathbf{R}|} + \frac{|\mathbf{r}|^2}{2a^2} -\frac{3}{2} \right) r^2\sin{\theta} dr d\theta d\phi \right\} \,.
 \label{int::R_le_2a}
\end{align}
\end{widetext}

\subsubsection{Case $a < R < a\sqrt{2}$}

From Figure \ref{fig::sph_int}(c) we see that:
\begin{equation}
 r_t^2+a^2 = R^2 \Longrightarrow r_t=\sqrt{R^2-a^2}
\end{equation}

\noindent and also
\begin{equation}
 \cos{\theta_t} = \frac{r_t}{R} = \frac{\sqrt{R^2-a^2}}{R} = \sqrt{1-\frac{a^2}{R^2}} \,.
\end{equation}

Then, for the integral we have:
\begin{widetext}
\begin{align}
 \int_{V_a-V_{\textrm{int}}} (c_\textrm{out}(\mathbf{r}) - c_\infty) d^3\mathbf{r} &= \displaystyle\frac{J}{4\pi D a}\left\{ \int_0^{2\pi}\int_0^{\theta_t}\int_0^{r_{-}} \left(\frac{a}{|\mathbf{r}-\mathbf{R}|} + \frac{|\mathbf{r}|^2}{2a^2} -\frac{3}{2} \right) r^2\sin{\theta} dr d\theta d\phi \ + \right. \notag \\
 & \hspace{2cm} \left. +\ \int_0^{2\pi}\int_{\theta_0}^{\theta_t}\int_{r_{+}}^{a} \left(\frac{a}{|\mathbf{r}-\mathbf{R}|} + \frac{|\mathbf{r}|^2}{2a^2} -\frac{3}{2} \right) r^2\sin{\theta} dr d\theta d\phi \right. \notag \\
 & \hspace{2cm} \left. +\ \int_0^{2\pi}\int_{\theta_t}^{\pi}\int_{0}^{a} \left(\frac{a}{|\mathbf{r}-\mathbf{R}|} + \frac{|\mathbf{r}|^2}{2a^2} -\frac{3}{2} \right) r^2\sin{\theta} dr d\theta d\phi \right\}\,,
 \label{int::R_le_asq2}
 \end{align}
 \end{widetext}
 
\noindent where $r_{\pm} = R\cos{\theta}\pm\sqrt{R^2\cos^2{\theta}-(R^2-a^2)}$.

 \subsubsection{Case $R < a$}
 
 For the case $R < a$ (Fig. \ref{fig::sph_int}(d)), we have:
 \begin{align}
  &R^2 + r_s^2 -2rR\cos{\theta} = a^2 \notag \\
  &r_s^2 -(2R\cos{\theta})r +(R^2-a^2) = 0 \notag \\
  &\Rightarrow r_s=R\cos{\theta}+\sqrt{R^2\cos^2{\theta}-(R^2-a^2)}
 \end{align}

 Thus, we have for the integral:
 \begin{widetext}
 \begin{align}
 \int_{V_a-V_{\textrm{int}}} (c_\textrm{out}(\mathbf{r}) - c_\infty) d^3\mathbf{r} =& \displaystyle\frac{J}{4\pi D a} \int_0^{2\pi}\int_{\theta_0}^{\pi}\int_{r_s}^{a} \left( \frac{a}{|\mathbf{r} - \mathbf{R}|} + \frac{|\mathbf{r}|^2}{2a^2} -\frac{3}{2} \right) r^2\sin{\theta} dr d\theta d\phi
 \label{int::R_le_a}
\end{align}
\end{widetext}

The integrals in Eqs.~(\ref{int::R_le_2a}), (\ref{int::R_le_asq2}) and (\ref{int::R_le_a}) yield the same result. Thus, the entry rate for $R<2a$, Eq.~(\ref{eq:alpha_gen}), is given by:
  \begin{align}
   \alpha &=\alpha_0c_{\infty}+\alpha_0\frac{J}{4\pi D a}\left[-\frac{R^2(-30a^2R+R^3+80a^3)}{160a^5}\right] \notag \\
   & = \alpha_0c_{\infty}+\alpha_0\frac{J}{4\pi D a}\left[\frac{3}{2}\left(\frac{R}{2a}\right)^3-\frac{1}{5}\left(\frac{R}{2a}\right)^5-2\left(\frac{R}{2a}\right)^2\right] \notag \\
   & = \alpha_0c_{\infty}+\alpha_0\frac{J}{4\pi D a}\left(\frac{R}{2a}\right)^2\left[\frac{3}{2}\left(\frac{R}{2a}\right)-\frac{1}{5}\left(\frac{R}{2a}\right)^3-2\right]\,.
  \end{align}

\section{Derivation of $l_\textrm{eff}$ \label{app:leff}}
The total unfolded length $L$ of an ideal polymer is given by $L=sN$, where $s$ is the Kuhn length and $N$ is the number of (non-interacting) monomers. 
From this model one can compute the mean of the squared end-to-end distance, i.e.
\begin{equation}
    \langle \mathbf{R}^2 \rangle = s^2 N\;.
\end{equation}

When considering a polymer composed of two different kinds of monomers, $N_1$ with Kuhn length $s_1$ and $N_2$ with Kuhn length $s_2$, one can write
\begin{equation}
    \langle \mathbf{R}^2 \rangle = s_1^2 N_1 + s_2^2 N_2\; \label{eq:R2}.
\end{equation}
Considering now the problem addressed in this work, $s_1 = 2 l_p$ corresponds to the Kuhn length of the polymer that is not covered by particles, while $s_2 = \ell$ is to the particle footprint. We constrain the total unfolded length of the polymer to be constant (the substrate does not change length if particles are on it): $L = 2 l_p N_1 + \ell N_2$. The number $N_1$ is given by $N_1 =\cfrac{L}{2 l_p} - \cfrac{n \ell}{2 l_p}$, while $N_2 = n$, where $n = \rho L$ is the total number of particles on the lattice. Plugging those relations into Eq.(\ref{eq:R2}) we obtain
\begin{equation}
    \sqrt{\langle \mathbf{R}^2 \rangle}= \sqrt{2L} \left[ l_p(1-\rho\ell) + \rho \frac{\ell^2}{2}\right]^{\frac{1}{2}}, 
\end{equation}
hence the definition of $l_\textrm{eff} = l_p(1-\rho\ell) + \rho \frac{\ell^2}{2}$ used in the main text.\\

\section{Parameters related to mRNA translation\label{app:mRNA}}
In the main text we mention that the parameters used in Fig.3 can roughly represent the translation of a typical mRNA. Due to the length constrain of the main text, we explain here the choice of the parameters used.

The typical length $L$ in codons of a gene is $\sim 300$, and the size of the ribosome footprint $\ell$ is around $10$ codons. Moreover, the persistence length of a mRNA is $\lesssim$ 1 codon. The radius of the reaction volume $a$ is related to the size of the ribosomes and to the 5' untranslated region (5'UTR) upstream the START codon (that could be though of as a landing platform for the ribosome). Parameter $a$ is then assumed to be between 1-2 ribosome sizes. It will also get larger according to the size of the untranslated region upstream the START codon. We then rescaled all the distances by $\ell$, obtaining the parameters of the order of magnitude used to plot Fig.3.

We are aware that this rescaling procedure does not reproduce exact rescaled elongation dynamics (since now each ribosome step corresponds to translocation through $\ell$ codons per step). However, this approximation shows that the effects of our framework could be important when considering biologically reasonable parameters, while still keeping the system algebraically and computationally tractable using the standard TASEP. More refined quantitative comparisons to biological systems will be considered when extending the model to the $\ell$-TASEP~\cite{Fernandes2018InPreparation}.

\section{Lattice simulations - Coupling polymer conformation and transport \label{app:sim}}

A simulation scheme can be introduced to obtain the  physical quantities of interest in a system where polymer conformation and transport are coupled, as we vary the size $L$ of the lattice. We obtain this with the following self-consistent method:
\begin{itemize}
\item[(i)] We initialise the system with an arbitrary small value of $\alpha = \alpha^{(init)}$, let the system evolve until the steady state is reached and then evaluate the current $J^{(init)}$ and the density $\rho^{(init)}$;

\item[(ii)] Compute $R$ as in equation (9) and update $\alpha = \alpha^{(fin)}$ according to equation (5), with $\Gamma$ given by equation (12), and current $J^{(init)}$ and density $\rho^{(init)}$ computed in (i);

\item[(iii)] If $|\alpha^{(fin)}-\alpha^{(init)}|/\alpha^{(fin)}<0.01$, stop the iteration. If not, give a small increment to $\alpha^{(init)}$ and repeat steps (i) and (ii) until convergence;

\item[(iv)] The converging value of $\alpha$ is then used to obtain the final densities and currents.
\end{itemize}

Thus, for a given choice of the parameters (see Fig.3 in the main text), we can obtain the steady state-density $\rho$ and initiation rate $\alpha$, which vary with the length $L$ of the lattice, considering the contributions of transport (here including particle recycling and finite-size effects) and the conformational state of the polymer.

Since the agreement with the model is excellent, this simulation scheme is also a proof that the finite size effects, which are intrinsic to the numerical simulations, are not a dominating process.

\section{Phase diagrams with self-avoiding polymer model \label{app:SAW}}
\begin{figure}[bh]
\centering
\includegraphics[width = 0.9\columnwidth]{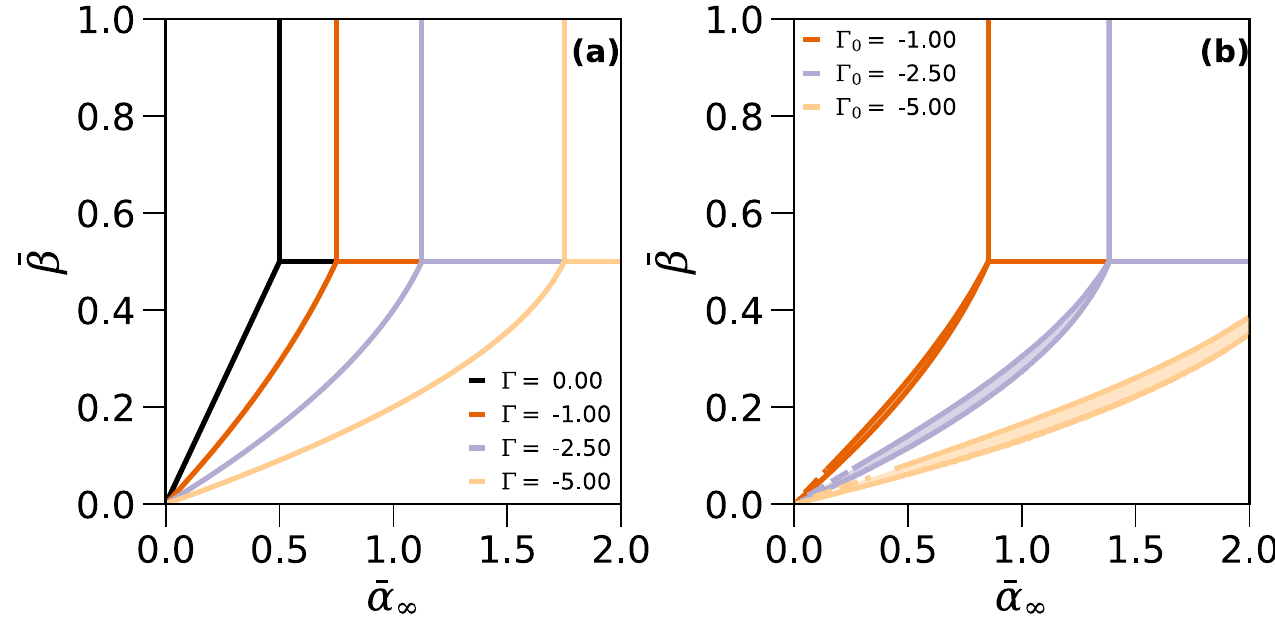}
\caption{(Color online) Phase diagrams (equivalent of Fig.2 of the main text) computed with a self-avoiding polymer model $R\propto L^{0.588}$.
}
\label{fig::self-avoiding}
\end{figure}

Throughout the text we used an ideal polymer model with $R \propto \sqrt L$. We decided to use that relation for the sake of simplicity, bearing in mind that other choices of how $R$ scales with $L$ should be used when trying to quantitatively compare this model to experimental data. However, the physics behind the phenomenology does not change with other choices, as we show in the Figure~\ref{fig::self-avoiding} (equivalent to Fig.2 of the main text) where we used a self-avoiding polymer $R\propto L^{0.588}$.

\bibliographystyle{apsrev4-1}
\bibliography{biblio_polymerTASEP.bib}
\end{document}